\documentclass[sn-mathphys-num,Numbered]{sn-jnl}

\usepackage{graphicx}%
\usepackage{multirow}%
\usepackage{amsmath,amssymb,amsfonts}%
\usepackage{amsthm}%
\usepackage{mathrsfs}%
\usepackage[title]{appendix}%
\usepackage{xcolor}%
\usepackage{textcomp}%
\usepackage{manyfoot}%
\usepackage{booktabs}%
\usepackage{algorithm}%
\usepackage{algorithmicx}%
\usepackage{algpseudocode}%
\usepackage{listings}%

\begin{document}

\title[Free-Space Twin-Field Quantum Key Distribution]{Free-Space Twin-Field Quantum Key Distribution}


\author[1,2,3]{\fnm{Yu-Huai} \sur{Li}}
\equalcont{These authors contributed equally to this work.}

\author[1,2,3]{\fnm{Ting} \sur{Zeng}}
\equalcont{These authors contributed equally to this work.}

\author[1,2,3]{\fnm{Min-Yan} \sur{Wang}}
\equalcont{These authors contributed equally to this work.}

\author[4]{\fnm{Cong} \sur{Jiang}}

\author[1,2,3]{\fnm{Jin} \sur{Lin}}

\author[1,2,3]{\fnm{Hao-Bin} \sur{Fu}}

\author[1,2,3]{\fnm{Xin-Yang} \sur{Zheng}}

\author[1,2,3]{\fnm{Jiu-Peng} \sur{Chen}}

\author[6]{\fnm{Zeng-Sen} \sur{Lin}}

\author[1,2,3]{\fnm{Cheng-Lin} \sur{Li}}

\author[1,2,3]{\fnm{Jian-Yu} \sur{Guan}}

\author[1,2,3]{\fnm{Yang} \sur{Li}}

\author[1,2,3]{\fnm{Qi} \sur{Shen}}

\author[5]{\fnm{Hao} \sur{Li}}

\author[5]{\fnm{Lixing} \sur{You}}

\author[5]{\fnm{Zhen} \sur{Wang}}

\author[3,6]{\fnm{Fei} \sur{Zhou}}

\author[1,2,3]{\fnm{Juan} \sur{Yin}}

\author[1,2,3]{\fnm{Sheng-Kai} \sur{Liao}}

\author[1,2,3]{\fnm{Ji-Gang} \sur{Ren}}

\author[2,3,4]{\fnm{Xiang-Bin} \sur{Wang}}

\author[1,2,3]{\fnm{Yuan} \sur{Cao}}

\author[1,2,3]{\fnm{Qiang} \sur{Zhang}}

\author[1,2,3]{\fnm{Cheng-Zhi} \sur{Peng}}

\author[1,2,3]{\fnm{Jian-Wei} \sur{Pan}}

\affil[1]{\orgdiv{Hefei National Research Center for Physical Sciences at the Microscale and School of Physical Sciences}, \orgname{University of Science and Technology of China}, \orgaddress{\city{Hefei}, \postcode{230026}, \country{China}}}

\affil[2]{\orgdiv{Shanghai Research Center for Quantum Sciences and CAS Center for Excellence in Quantum Information and Quantum Physics}, \orgname{University of Science and Technology of China}, \orgaddress{\city{Shanghai}, \postcode{201315}, \country{China}}}

\affil[3]{\orgdiv{Hefei National Laboratory}, \orgname{University of Science and Technology of China}, \orgaddress{\city{Hefei}, \postcode{230088}, \country{China}}}

\affil[4]{\orgdiv{State Key Laboratory of Low Dimensional Quantum Physics}, \orgname{Tsinghua University}, \orgaddress{\city{Beijing}, \postcode{100084}, \country{China}}}

\affil[5]{\orgdiv{State Key Laboratory of Functional Materials for Informatics}, \orgname{Shanghai Institute of Microsystem and Information Technology, Chinese Academy of Sciences}, \orgaddress{\city{Shanghai}, \postcode{200050}, \country{China}}}

\affil[6]{\orgname{Jinan Institute of Quantum Technology}, \orgaddress{\city{Jinan}, \postcode{250101}, \country{China}}}

\abstract{
Twin-field quantum key distribution (TF-QKD) elevates the secure key rate from a linear to a square-root dependence on channel loss while preserving measurement-device-independent security. This protocol is uniquely positioned to enable global-scale quantum networks, even under extreme channel loss. While fiber-based TF-QKD implementations have advanced rapidly since its proposal, free-space realizations have remained elusive due to atmospheric turbulence-induced phase distortions. Here, we report the first experimental demonstration of free-space TF-QKD over 14.2 km urban atmospheric channels, surpassing the effective atmospheric thickness—a critical threshold for satellite compatibility. We achieve a secret key rate exceeding the repeaterless capacity bound, a milestone for practical quantum communication. Our approach eliminates the need for an auxiliary channel to stabilize a closed interferometer, instead leveraging open-channel time and phase control of optical pulses. This work represents a pivotal advance toward satellite-based global quantum networks, combining high-speed key distribution with inherent resistance to real-world channel fluctuations.
}

\maketitle

In principle, quantum key distribution (QKD) offers information-theoretic security for private communication based on the laws of physics \cite{BB84, E91, RevModPhys.74.145, RevModPhys.81.1301}.
After decades of development, the most important issues currently in the QKD field are the large-scale coverage and the security with realistic devices.
For the former, a series of experiments based on Micius quantum science satellite \cite{liao2017satellite, Liao2018relay, yin2017satellite, Yin2020, Yin2017entQKD, ren2017satellite, Chen2021, RevModPhys.94.035001} were implemented, demonstrating the feasibility of realizing the global-scale QKD with satellite platforms.
For the latter, although attacks exist in practical QKD, many efforts have been made toward the goal of secure communication with imperfect devices. 
After numerous attempts, researchers now thoroughly understand and can manage the practical imperfections~\cite{RevModPhys.92.025002}. 
For example, the protocol of measurement-device-independent QKD (MDI-QKD) \cite{LoMDI2012, Braunstein2012sidefree} guarantees the security of QKD against any detector imperfections. 
The MDI-QKD combined with the decoy-state method \cite{Wang:Decoy:2005, Lo:Decoy:2005} has become a promising approach for securing QKD with realistic devices.

However, obtaining a meaningful secure key rate with large-scale channels is still challenging owing to the high loss.
Twin-field QKD (TF-QKD) \cite{Lucamarini2018overcoming}, which improves the key rate scaling to follow the square root of the channel transmittance, is an inspiring solution for the global-scale quantum communication network.
More importantly, the characteristic of MDI is maintained in the TF-QKD protocol, thus the practical security is not lowered.
Great efforts have been made to demonstrate TF-QKD with the fiber channel \cite{mindere2019, PhysRevX.9.021046, PhysRevLett.123.100505, PhysRevLett.123.100506, fang_implementation_2020, PhysRevLett.124.070501, PhysRevLett.126.250502, chen_twin-field_2021, PhysRevLett.128.180502, pittaluga_600-km_2021, wang_twin-field_2022, PhysRevLett.130.210801, zhou_twin-field_2023, PhysRevLett.130.250801, PhysRevLett.130.250802, PhysRevLett.132.260802} through up to 1002-km spooled fiber \cite{PhysRevLett.130.210801} and in the field test through 511-km deployed fiber between metropolitans \cite{chen_twin-field_2021}.

Extending the TF-QKD into the free-space channel is still challenging.
The intense atmosphere turbulence causes intensity scintillations and phase discontinuity on the arrival beams, making it difficult to guarantee the indistinguishability between photons propagated through different free-space channels.
Moreover, suppressing phase noise between two independent lasers is challenging due to the lack of stable channels, which is also a more difficult aspect compared to free-space MDI-QKD~\cite{Cao2020, LiMDI2023}.
Here, we overcome these challenges by employing two ultra-stable lasers (USL) with extremely low phase noise and two Rb atomic clocks in Alice and Bob as the photon source and the time reference.
The frequencies of the USLs and the atomic clocks are feedbacked by the statistics of remotely detected single photon events, thus insensitive to the atmospheric turbulence and avoiding additional locking laser needs.
We finally perform an experimental demonstration of TF-QKD over 14.2 km with turbulence free-space channels, with a secure key rate exceeding the Pirandola-Laurenza-Ottaviani-Bianchi (PLOB) bound \cite{pirandola_fundamental_2017} corresponding to the same channel loss.
The distance of the free-space channel in our work is comparable to the effective thickness of the atmosphere ($\sim10~km$) \cite{RevModPhys.94.035001}, and the turbulence of the horizontal atmospheric channel is normally much stronger than that of the vertical channel. Thus, such a demonstration has presented a significant step towards satellite-based TF-QKD.

Furthermore, the realization of single-photon interference over long-distance free-space channels in this work opens up the possibilities of performing complex quantum information tasks at a large scale, such as quantum teleportation \cite{Teleportation1993}, quantum repeaters \cite{Briegel1998repeater}, distributed quantum computation \cite{PhysRevA.59.4249}, or even testing quantum effects in curved spacetime \cite{rideout_fundamental_2012}.

\begin{figure}[!t]\center
\includegraphics[width=1\textwidth]{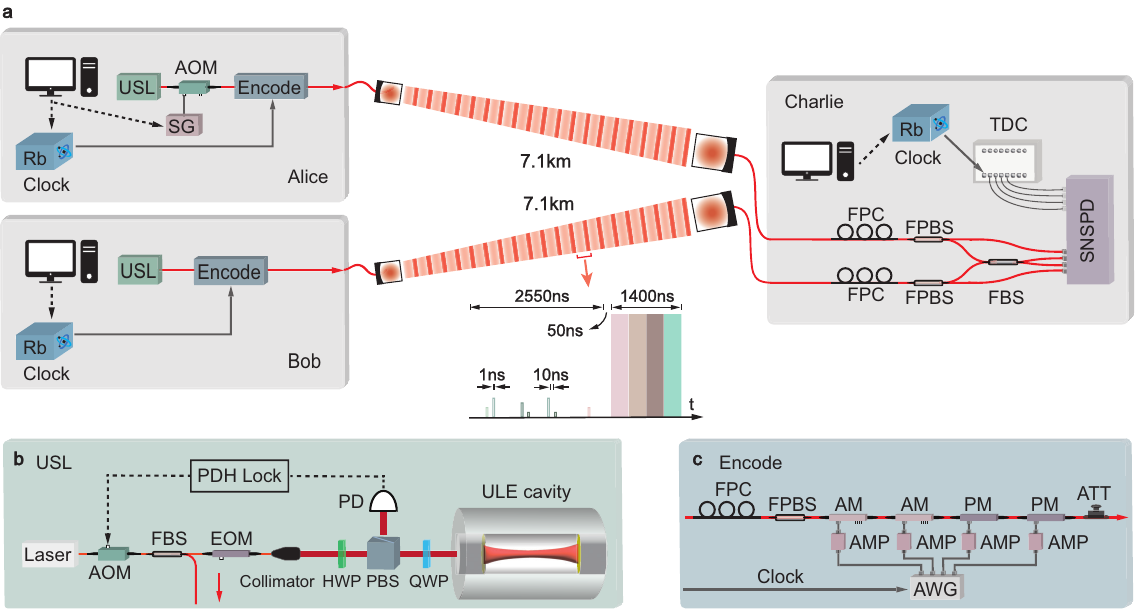}
\caption{Setup of the free-space TF-QKD. The senders and the receiver are connected by two free-space channels over the urban atmosphere in Shanghai, with a total distance of $14.2~km$. (a) A CW ultra-stable laser (USL) is employed in each sender as the photon source. The wavelength in Alice can be slightly tuned by an acoustic-optic modulator (AOM) to compensate for the slow drift. After being modulated in the encoding module, photons are sent to the atmospheric channel by a telescope with a diameter of $150~mm$. Received optical pulses interfere at Charlie's fiber beam splitter (FBS), measured by the superconducting nanowire single-photon detectors (SNSPDs) with a detection efficiency of approximately 80\% and dark counts below 20 cps, and recorded on a time-digital converter (TDC). Before the interference, a fiber polarization controller (FPC) and a fiber polarizing beam splitter (FPBS) are employed for each channel to ensure polarization indistinguishability and separate a portion of photons for the arrival time measurement. Rb clocks with one second Allan Variance better than $2\times10^{-11}$ are employed as the time reference for the TDC and the encoding modules. The clocks are calibrated periodically by the pulse arrival time. (b) The USL is constructed by locking a seed laser to an ultra-low-expansion (ULE) cavity with the Pound-Drever-Hall (PDH) technology for good frequency stability. (c) The detail of the encoding module. Pulses, intensities, and phases are modulated on the laser by two amplitude modulators and two phase modulators. These modulators are driven by an arbitrary waveform generator (AWG). EOM, electro-optic modulator; PBS, polarizing beam splitter; HWP, half-wave plate; QWP, quarter-wave plate; PD, photodiode; ATT, attenuator; AMP, amplifier.}
\label{Fig:Setup}
\end{figure}

In this experiment, Alice and Bob are connected with Charlie with two $7.1~km$ free-space channels, as shown in Fig. \ref{Fig:Setup}.
As single-photon interference happens for the X-basis~\cite{PhysRevA.98.062323}, the quantum bit error rate in the X-basis ($QBER_{X}$) in TF-QKD is sensitive to the relative phase $\Delta\varphi$ between arrival coherence pulses.
Therefore, the realization of TF-QKD is challenging compared to other QKD protocols due to the requirement of precise phase control from independent lasers through long-distance free-space channels for high visibility of interference.
Suppose the frequencies of photons emitted from Alice and Bob are $\nu$ and $\nu+\Delta\nu+\delta\nu$ respectively.
Here, $\delta\nu$ is the unexpected frequency deviation introduced by the laser and the atmospheric channel.
$\Delta\nu \ll \nu$ is the preset frequency difference for the heterodyne detection. To obtain a clear beating signal, $\Delta\nu$ is also required to be observably larger than $\delta\nu$.
When estimating $\Delta\varphi$ with photons in a duration of $\tau$, the estimating error $\delta\varphi$ can be approximatively regarded as
\begin{equation}
\label{dphi}
    \delta\varphi = 2\pi\tau\delta\nu + 2\pi\nu\frac{\Delta L(\tau)}{c}
\end{equation}
where $\Delta L(\tau)$ is the fluctuation of the channel, and $c$ is the speed of light.
The two terms in Formula \ref{dphi} represent the main sources of phase noise, the frequency deviation of the laser, and the fluctuation of the channel.

The key challenge here is determining $\Delta\varphi$ at each moment with a sufficiently small error ($\delta\varphi$). This is achieved in three stages: suppressing the short-term frequency stability and phase noise of the laser, feedback compensating for long-term frequency drift of the laser, and calculating the relative phase $\Delta\varphi$ by single photon detection events.

As the first stage, previous fiber-based experiments involved additional fiber channels for the locking of lasers between Alice and Bob to suppress the short-term phase noise.
In contrast, atmospheric turbulence inevitably affects the wavefront of laser beams and results in a varying random distribution of amplitude and phase at the receiving aperture, leading to a fluctuating intensity in free space.
Hence, it is challenging to perform such a phase locking between independent lasers via a free-space channel.
Therefore, to suppress the phase noise of the laser, a USL is installed in each of the encoding terminals as the photon source.
A continuous wave laser with a center wavelength of $1550.12~nm$ and a line width of several kilohertz is used as the seed laser.
Such seed laser is locked to a $10~cm$-long ultra-stable cavity with a finesse of around $250,000$ to suppress the line width to a few hertz \cite{PhysRevLett.93.250602} by the Pound-Drever-Hall (PDH) technique \cite{drever1983laser, pound1946electronic}, as shown in Fig. \ref{Fig:Setup}(b).
With the PDH locking, the frequency characteristics of the output laser are mainly determined by the stability of the cavity length.
The shape and supporting system of the cavity is specialized to minimize the influence of vibration.
To suppress the long-term drift, the cavity is placed in a chamber with two thermal shields.
The chamber also maintains a high vacuum to suppress thermal convection as well as the influence of the airflow.
The long-term frequency drift is approximately $0.15~Hz/s$ with quite low short-term fluctuation.
Thus, it is sufficient for measurement and feedback in the time scale of several seconds over the free-space channel.

\begin{figure}
\centering
\includegraphics[width=0.8\textwidth]{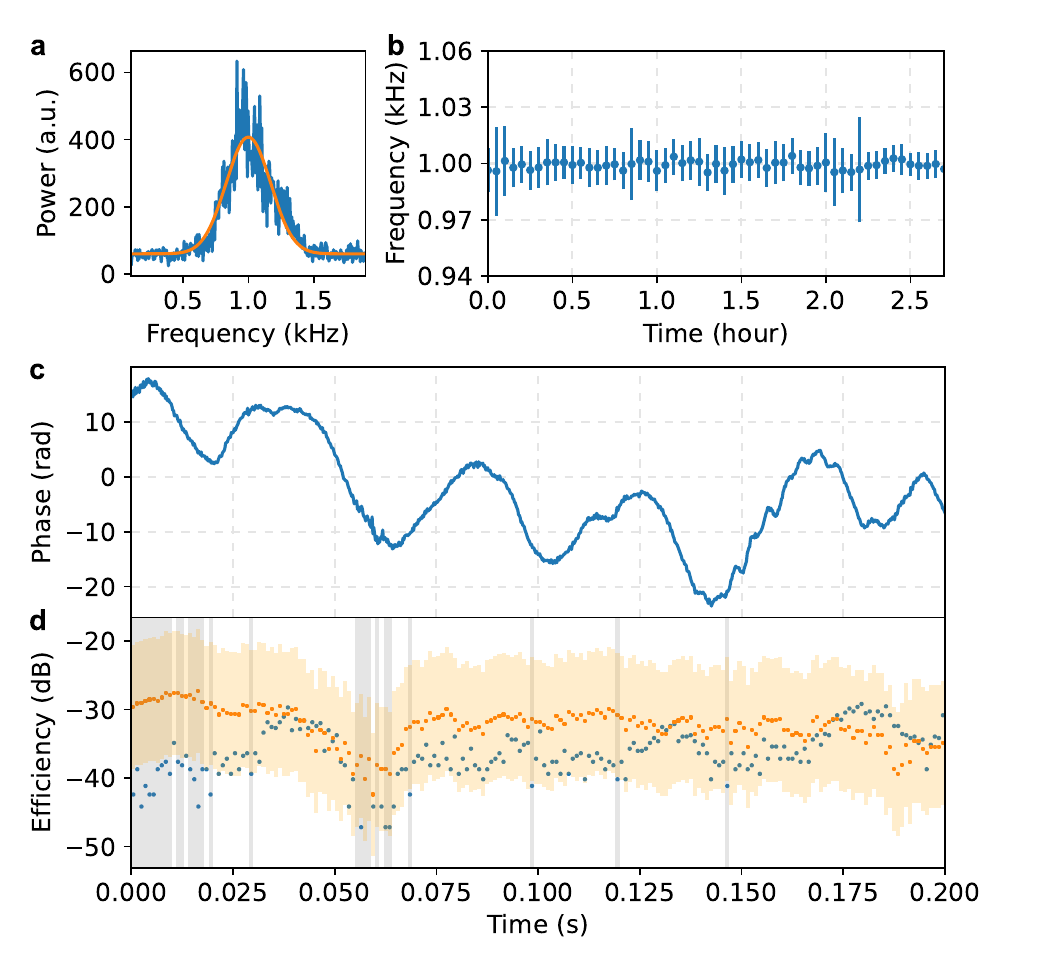}
\caption{\label{fig:noise} (a) The spectral span of ultra-stable laser beating after propagating the $7.1~km$ atmospheric channel. (b) With a real-time calibration, the central wavelength difference of the two lasers is fixed at $1~kHz$, with an overall deviation of $\sigma=10.9~Hz$. (c) Typical phase drift in $200~ms$. (d) Typical channel efficiency variation in $200~ms$. Blue (orange) markers denote the instantaneous channel efficiency from Alice (Bob) to Charlie. For any instant, the blue marker falls inside (outside) the yellow shadow means the efficiency ratio between the two channels is smaller (larger) than a specified threshold. In certain duration (gray shadowed), a significant efficiency mismatch appears due to the strong turbulence of the atmosphere, leading to low interference visibility. Thus, the corresponding data has to be discarded.}
\end{figure}

The goal of the second stage further compensates for the residual frequency drift in the long term. The intense coherent pulses emitted from Alice and Bob interfere with a fiber beam splitter (FBS) in Charlie.
By statistic the single photon detection rate in the two SNSPD channels and converting to the frequency domain, the frequency difference and span of the arrival photons can be obtained, as shown in Fig. \ref{fig:noise}(a).
As the frequency broadening is below $\sim 300~Hz$ for the atmospheric channel transmitted photons, we set the center frequencies of the two USLs to be separated by $\Delta\nu=1~kHz$ for continuous calibration.
An acoustic-optic modulator (AOM) is installed in Alice for the calibration.
The distribution of center frequency differences after the periodic calibration is shown in Fig. \ref{fig:noise}(b), with a deviation of $10.9~Hz$. 

Once the frequency difference between the photon sources is determined and locked, the third stage is performed to estimate the relative phase $\Delta\varphi$ by interference measurement.
The typical phase drift during $200~ms$ is shown in Fig. \ref{fig:noise}(c), which contains the phase noise from the free-space channels and the residual phase noise of the laser.
In certain moments, a significant efficiency mismatch leads to a low visibility of interference and an unreliable estimated phase.
With such durations (as gray shadowed in Fig. \ref{fig:noise}(d)) been excluded, the standard deviation of the phase drift rate is about $2.1~rad/ms$, roughly equivalent to the noise level of $\sim 100~km$ fiber \cite{Lucamarini2018overcoming, PhysRevLett.123.100505}.
It is worth noting that, the distance of free-space channel in the current setup is comparable to the effective thickness of the atmosphere
\cite{RevModPhys.94.035001}.
Therefore, it is expected that the phase noise induced by the atmosphere will not increase significantly when performing satellite-based TF-QKD, with the channel distance over hundreds or even thousands of kilometers.

Each of the free-space channels is established by a $127~mm$ diameter telescope as the sender and a $400~mm$ diameter telescope as the receiver.
The arrival photons are coupled into a single-mode fiber (SMF) in each of the receivers.
The SMF coupling efficiency fluctuates dramatically along with the turbulence of the atmosphere, which is the main source of attenuation in the free-space channel and highly dependent on the strength of atmospheric turbulence described by Fried's coherence length $r_0$ \cite{Fried:66}.
To obtain sufficient visibility of twin-field interference, indistinguishability in several freedoms of degree is required to be ensured.
The received photons from each channel are split into two portions by a fiber polarizing beam splitter (FPBS).
By controlling the polarization of input beams, approximately $10\%$ of the photons are transmitted through the FPBS for the interference measurement, while the remaining serves as the probe.
As a polarizing filtering, the FPBS ensures the output photons form identical polarization before interference.
Slight polarization deviation, mainly introduced by the fiber stress variation, will be converted to intensity deviation.
Therefore, the influence on the interference visibility is significantly reduced.
All the arrival photons are detected by 4 channels of SNSPDs, as shown in Fig. \ref{Fig:Setup}.
As the probe photons are detected without mixing with photons from another channel, they can exhibit properties of the current channel, especially the time and efficiency.
The arrival times, compared to Charlie's clock, reflect the time variation between Alice (Bob) and Charlie.
Together with Rb atomic clocks employed in each of the three terminals as the time reference, the time synchronization can be realized by performing the time drift measurement and calibration every several seconds.
The single-photon detection event counting rate during a certain duration is another good indicator that reflects the instantaneous efficiency of the current channel.
Although the SMF coupling is the most significant source of loss in the setup, it is actually irreplaceable as a necessary burden to ensure indistinguishability in the spatial mode.
By the SMF coupling, spatial mode variations turn into efficiency variations, which are much easier to measure and quantify.
Therefore, data acquired in durations with significantly different efficiencies between the two free-space channels can be discarded to increase the interference visibility effectively.
The efficiency estimation duration is $1~ms$ in this experiment, which is faster than typical fluctuations of atmospheric turbulence.
It is worth noting that although Charlie plays a key role in frequency and time calibration and phase and efficiency estimation, the security of our implementation can still be ensured without relying on trust in Charlie due to the measurement-device-independent nature of the proposal.
Any deviation, vandalism, or eavesdropping by Charlie, as well as by any adversaries in the channel, can only reduce the secure key rate rather than the security of obtained keys.

Following the four-intensity decoy state sending-or-not-sending protocol \cite{PhysRevA.98.062323}, Alice and Bob randomly generate coherent optical pulses with one of the intensities $\mu_z$, $\mu_x$, $\mu_y$, and $0$ by two amplitude modulators (AMs).
For each pulse in Alice (Bob), a random phase encoding $e^{i\varphi_A}$ ($e^{i\varphi_B}$), where $\varphi_A, \varphi_B = k\pi/8$ ($k=0, 1, ..., 15$), is then prepared by two phase modulators (PMs).
After that, the pulses are transmitted to the measurement terminal Charlie, superposed on a beam splitter (BS), and detected by superconducting nanowire single-photon detectors (SNSPDs).
An X-basis event is regarded as effective if Charlie announces only one detector clicked, and the phase encoding satisfies $1-|cos(\varphi_A - \varphi_B + \Delta\varphi(t))| < \Lambda$ \cite{PhysRevLett.123.100505}.
Here, $\Lambda$ is the limit presenting the tolerance of $QBER_{X}$, which can be determined afterward.
$\Delta\varphi_t$ is the phase noise at time $t$, which is mainly introduced by the photon source and the channel.
As the major sources of phase noises are characterized, the influence on the $QBER_{X}$ can be restricted by choosing a proper phase difference estimating duration.
A longer duration allows more reference photons for a more accurate phase estimation but leads to a higher phase variation at the same time, thus needing a tradeoff.
In the post-process of our experiment, the estimation duration is set to be $200~\mu s$.

Similar to the previous experimental works, strong reference pulses are inserted into the stream of quantum signal pulses to estimate the phase difference.
As the timing sequence in Fig. \ref{Fig:Setup} shows, the strong reference pulses and the quantum signal pulses are emitted alternately within a period of $4~\mu s$, denoted as a \textit{frame}.
A $50~ns$ gap presents after the reference pulses, allowing the dark count of the SNSPDs to recover from the strong illumination.
Details about the phase estimation process can be found in the Methods.
Since the phase difference can be accumulated by any unwanted wavelength differences between the photon sources or by the fast phase drift in the free-space channel, the residual error after phase estimation should be limited.

\begin{figure}
\centering
\includegraphics[width=0.76\textwidth]{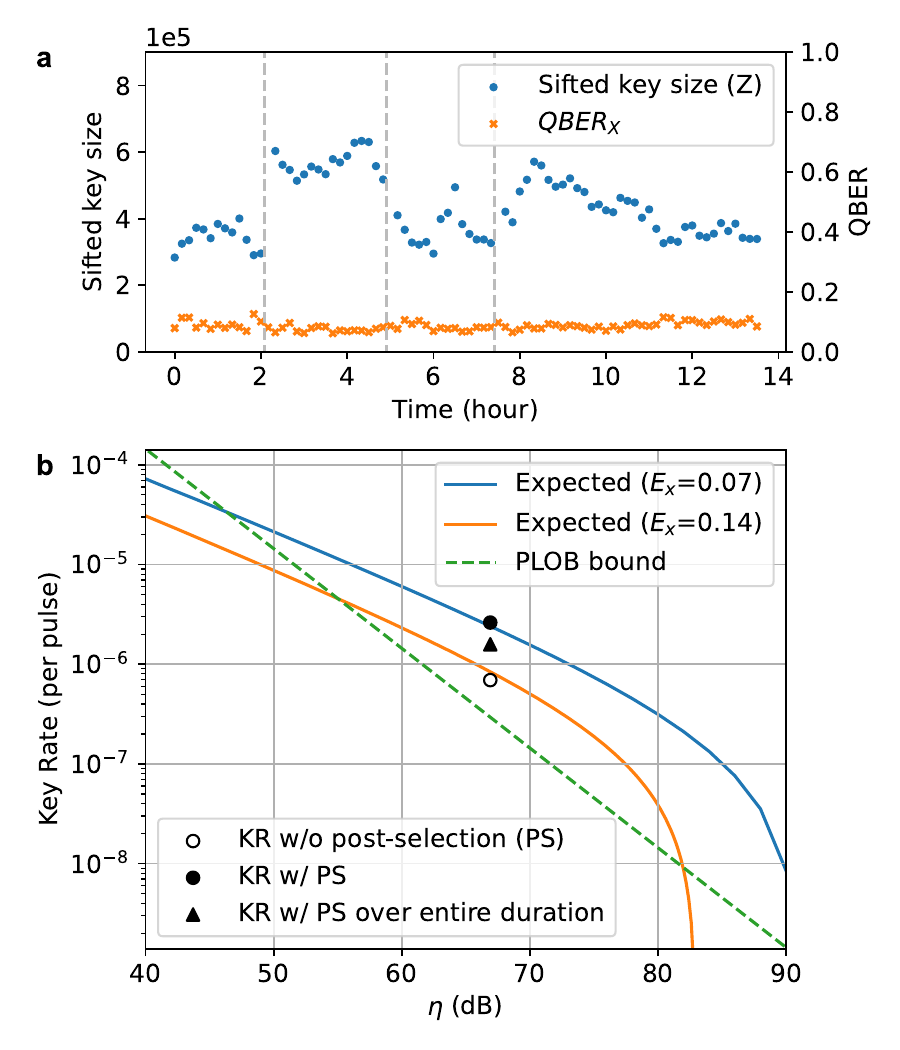}
\caption{\label{fig:result} Results of free-space TF-QKD. (a) The sifted key rate in Z basis and QBER in X basis over the whole data acquiring duration grouped every $10~minutes$. The $QBER_X$ is calculated after the efficiency post-selection with a threshold of $0.23$, with an average of $7.36\%$. (b) With the total valid time of $49,359~s$, the secure key rate is $6.92\times10^{-7}$ per pulse (hollow circle), which is close to the expected curve with $QBER_X=0.14$ (orange curve). After the post-selection process, $29,859~s$ of valid time is selected with the secure key rate of $2.62\times10^{-6}$ (solid circle), which is close to the simulated curve with $QBER_X=0.07$ (blue curve). The triangle marker represents the effective secure key rate of post-selected data over the entire data-taking duration, i.e., $49,359~s$, which is the practically meaningful key rate. The above key rates have exceeded the PLOB bound (dashed curve). All the estimations of secure key rates have considered the finite key length.}
\end{figure}

The experimental data was taken for $49,359~s$ of valid time.
With $100~MHz$ of pulse repetition rate and $63.75\%$ of quantum signal pulse proportion in each \textit{frame}, approximately $3.15\times10^{12}$ quantum signal pulses were sent, and $54,404,847$ sifted keys in Z basis were obtained.
The source parameters are summarized in the Methods, which were optimized beforehand according to the setup.
The overall average QBER in Z and X bases are $28.1\%$ ($0.202\%$ after actively odd parity pairing (AOPP) \cite{Jiang_2020}) and $13.8\%$, respectively.
The raw key sizes and QBERs for every $10$ minutes are shown as Fig. \ref{fig:result}(a).
With the consideration of finite key size, $2.178~Mbits$ of secure keys are obtained in total with all the acquired data, corresponding to a key rate of $44.1$ bps.
Here, the existence of atmospheric turbulence dramatically disturbs the channel efficiency, leading to a relatively high $QBER_X$ that lowers the secure key rate.
By properly post-selecting the acquired data with the counting rate of probe photons indicating the instantaneous efficiencies, a higher secure key rate is expected, although some of the data will be discarded.
As shown in Fig. \ref{fig:result}(b), by post-selecting the data with efficiency ratio between $r_p$ and $1/r_p$, where the threshold $r_p=0.23$, $1.90\times10^{12}$ valid pulses are selected with the secure key of $2.61\times10^{-6}$ per pulse considering the finite key size.
Therefore, the total secure key size is increased to $4.980~Mbits$, corresponding to a key rate of $100.8$ bps.
The average efficiency from Alice and Bob to Charlie is approximately $67~dB$ in total.
As shown in Fig. \ref{fig:result}(b), the secure key rates above have exceeded the corresponding PLOB bound.

In conclusion, we have, for the first time, experimentally demonstrated TF-QKD over long-distance free-space channels.
We noticed that additional fiber is also eliminated by other approaches, like using frequency comb or locking the laser frequency to the absorption spectroscopy of molecules in recent fiber-based implementations \cite{zhou_twin-field_2023, PhysRevLett.132.260802}, which can also work in the free space TF-QKD.
Such a technical makes TF QKD more practical for various scenarios, especially the satellite-based situation.
As a novel protocol that can have a secure key generation rate exceeding the repeaterless bound and measurement-device-independent security, TF-QKD has great potential in the global quantum communication network.
With the assistance of an (untrusted) satellite, QKD can be performed between distant parties, even on opposite sides of the earth, with extremely high loss.
Surely, there are still many technical challenges that exist for realizing satellite-based TF-QKD, including the frequency and time calibration under the Doppler frequency shift induced by the fast-moving satellite and spaceborne single-photon detectors \cite{You:s}.
Overcoming these challenges will be the next goal for us.

Moreover, when realizing this protocol, a huge single-photon interferometer, with the length of $\sim 10~km$ for each arm, was constructed. Such an interferometer is an invaluable tool for many other tasks, such as quantum repeaters \cite{PhysRevLett.81.5932}, quantum network \cite{halder_entangling_2007}, geodesy \cite{anderson_sagnac_1994}, or even measuring the gravitational redshift and time dilation \cite{Rideout_2012}.

\section*{Acknowledgments}\label{Sec.Ackn}

We acknowledge insightful discussions with Wei Li and Yang Liu. We acknowledge open source library \textit{InteractionFree} \cite{interactionfree}, which enabled us to easily build the distributed system. This work has been supported by the National Key R$\&$D Program of China (Grants No. 2020YFA0309803, No. 2020YFC2200103), the National Natural Science Foundation of China (Grants No. 12174374, No. 12274398, No. 12374475), the Innovation Program for Quantum Science and Technology (Grants No. 2021ZD0300100), the Chinese Academy of Sciences (CAS), Shanghai Municipal Science and Technology Major Project (Grant No. 2019SHZDZX01), the Shanghai Rising-Star Program (Grants No. 21QA1409600), Key R$\&$D Plan of Shandong Province (Grant No. 2020CXGC010105), CAS Young Interdisciplinary Innovation Team (Grants No. JCTD-2022-20), the CAS Project for Young Scientists in Basic Research (Grant No. YSBR-046, No. YSBR-085), and Anhui Initiative in Quantum Information Technologies. F. Zhou was supported by the Taishan Scholars Program, Shandong Provincial Natural Science Foundation (Grant No. ZR2023LLZ007). Y. Cao, Q. Shen, and Y.-H. Li was supported by the Youth Innovation Promotion Association of CAS (under Grants No. 2018492, No. 2021457, No. 2023475).

\section*{Methods}

\subsection*{M1. The four-intensity sending-or-not-sending twin-field QKD protocol.}

In the 4-intensity SNS protocol~\cite{PhysRevA.98.062323} applied here, there are four sources on Alice's (Bob's) side.
The source parameters are symmetric for Alice and Bob, i.e., Alice and Bob have the same intensity and emission probability settings.
The four sources are the vacuum source $v$ and the weak coherent state sources $x,y,z$.
Their intensities and probabilities are denote as $\mu_v=0,\mu_x,\mu_y,\mu_z$ and $p_v,p_x,p_y,p_z$ respectively.
In each time window, Alice (Bob) randomly prepares and sends out a pulse to Charlie.
Charlie is assumed to measure the interference result of the incoming pulse pair and announce the measurement results to Alice and Bob.
If only one of the two detectors clicks, Alice and Bob would take it as an effective event.
After Alice and Bob send $N$ pulse pairs to Charlie and Charlie announces all measurement results, Alice and Bob perform the following data post-processing.

For clarification, we denote $lr$ as the two-pulse source while Alice chooses the source $l$ and Bob chooses the source $r$ for $l,r=v,x,y,z$.
In this work, the effective event of sources $lr~(l=v,y,z;r=v,y,z)$ are used to extract the secure final keys.
We denote the number of effective events from source $lr$ as $n_{lr}$.
In the data post-processing, Alice and Bob first announce the locations of the time windows that either Alice or Bob chooses the source $x$.
After this, Alice and Bob shall know the locations of the time windows while they use the sources $vx,xv,xx$, while keeping the locations of the time windows while they use the sources $lr~(l=v,y,z;r=v,y,z)$ secret (unannounced)~\cite{hu2022universal}.
For those unannounced times windows, Alice (Bob) takes it as bit $0$ ($1$) if she (he) chooses the source $v$, and takes it as bit $1$ ($0$) if she (he) chooses the source $y$ or $z$, and the bits of those corresponding effective event from the unannounced windows form the raw key strings $Z_A$ and $Z_B$ in Alice and Bob's sides respectively.
Alice and Bob first perform the AOPP process to the raw key strings and then perform the error correction process.
For the rejected pairs in the AOPP, Alice and Bob announce the sources they choose in each time window.
After this, Alice shall know the value of $n_{yv}$, Bob shall know the value of $n_{vy}$, and both of them shall know the value of $n_{vv}$~\cite{hu2022universal}.
Then Alice announces the value of $n_{yv}$ to Bob.
With those values of $n_{vv},n_{vx},n_{xv},n_{vy},n_{yv}$, Bob can perform the decoy state analysis to get the lower bound of the number of untagged bits after AOPP.
Also, according to the phase information of the effective events from source $xx$, Bob can get the upper bound of the phase flip error rate of untagged bits after AOPP.
Finally, Alice and Bob can calculate the secure key rate according to the formulas shown in Ref. \cite{hu2022universal}:
\begin{equation*}
R=\frac{1}{N}\{n_1[1-H(e_1^{{ph}})]-fn_t H(E_t)\}-\frac{1}{N}[2\log_2{\frac{2}{\varepsilon_{cor}}}+4\log_2{\frac{1}{\sqrt{2}\varepsilon_{PA}\hat{\varepsilon}}}\\+2\log_2(n_t-\sum_{l,r=y,z}n_{lr})],
\end{equation*}
where $R$ is the key rate of per sending-out pulse pair; $n_1$ is the lower bound of the number of survived untagged bits after AOPP and  $e_1^{{ph}}$ is the upper bound of the phase-flip error rate of those survived untagged bits after AOPP; $n_t$ is the number of survived bits after AOPP and $E_t$ is the corresponding bit-flip error rate in those survived bits; $f$ is the error correction inefficiency which we set to $f=1.16$; $H(x)=-x\log_2x-(1-x)\log_2(1-x)$ is the Shannon entropy; $\varepsilon_{cor}$ is the failure probability of error correction; $\varepsilon_{PA}$ is the failure probability of privacy amplification; $\hat{\varepsilon}$ is the coefficient while using the chain rules of smooth min- and max- entropy~\cite{vitanov2013chain}.

The source parameters employed in the experiment were optimized accordingly, to obtain a maximal expected final key rate.
The detailed source parameters and the optimizing conditions were listed in Tab. \ref{Tab:SourceParameters}.

\begin{table}[h]
\label{Tab:SourceParameters}
\caption{Source parameters for the optimization of the symmetric four-intensity sending-or-not-sending protocol \cite{PhysRevA.98.062323} for TF-QKD: dark count rate $d_0$, misalignment error probabilities of the ${X}$ basis $E_d^{X}$, channel efficiency between Alice (Bob) and Charlie $\eta_A$ ($\eta_B$), efficiency of the measurement module at Charlie $\eta_M$, error-correction efficiency $f$, failure probability in the statistical fluctuation analysis $\epsilon$, and total number of pulse pairs N. The optimized source parameters are listed in the lower part of the table.}\label{tab:source_para}
\begin{tabular}{c|ccccccc}
\hline
\multirow{2}{*}{Conditions} & $d_0$ & $E_d^{X}$ & $\eta_A$, $\eta_B$& $\eta_M$ & $f$ & $\epsilon$ & $N$ \\
\cline{2-8}
& $4\times10^{-8}$ & $5\%$ & $27~dB$ & $3~dB$ & $1.10$ & $10^{-10}$ & $10^{12}$ \\
\hline
Optimized & $\mu_{x}$ & $\mu_{y}$ & $\mu_z$ & $p_{x}$ & $p_{y}$ & $p_z$ & \\
\cline{2-8}
Parameters & $0.030$ & $0.483$ & $0.494$ & $0.063$ & $0.209$ & $0.053$ & \\
\hline
\end{tabular}
\end{table}

\subsection*{M2. Phase estimation.}
The fluctuating relative phases between photons emitted from two encoding terminals are required to be determined for each pulse pair to obtain a reference frame alignment.
Therefore, strong optical pulses that contain a large number of photons are employed for this task.
These strong pulses, named reference pulses, are emitted alternately with the quantum signal pulses from the same encoding terminals.
As reference pulses are required to be coherent with the signal pulses, these two kinds of pulses are generated by externally modulating the same cw laser by AMs.
They have the same optical frequency and continuous phase and experience the same transmission channel.
As shown in Fig. X, an encoding \textit{frame} has a total duration of $4,000~ns$, which consists of $1,400~ns$ of reference pulses, $2,550~ns$ of signal pulses, and $50~ns$ of gap.
The gap exists after the reference pulses for the SNSPD to recover to a low noise level.
All the pulses are modulated with a repetition rate of $100~MHz$.
For signal pulses, the modulation is determined by a stream of random numbers according to the SNS protocol, with the pulse width of $1~ns$, intensities randomly selected in $\{0,\mu_x,\mu_y,\mu_z\}$ and phases randomly selected from $k\pi/8$ where $k=0,1,...,15$.
For reference pulses, the pulse width is extended $10~ns$ with uniform strength.
Since the output intensity of an MZ-interferometer varies sinusoidally with phase, which is not a single-valued function, the reference pulses are modulated with four possible phases $0$, $\pi/2$, $\pi$, or $3\pi/2$.
Though not required for security, the applied phase for each reference pulse is also randomly picked to keep the PMs having a consistent performance between the reference pulses and the signal pulses.
After being detected by the SNSPDs in Charlie's terminal, the times of each detector's click are recorded by the TDC for further analysis.

Suppose $N$ detection events are recorded during a certain phase estimation duration, denoting $E(1), E(2), ..., E(N)$.
For detection event $E(i)$, the arrival time is $t(i)$, and the clicked detector channel is $c(i)$ ($c(i)=0$ or $1$), corresponding to the two output ports of the interferometer.
According to the arrival time $t(i)$, the modulated phase $\varphi_{mod}(i)$ and the phase introduced by the laser frequency difference $\varphi_{\nu}(i)=2\pi\Delta\nu t(i)$ can be determined.
To estimate the unknown phase $\varphi_c$, all the detection events are categorized into $M$ groups, denoting $G(1), G(2), ..., G(M)$.
The $i$th event $E(i)$ is placed in group $G(g(i))$, where $g(i)=\lfloor\frac{M}{2\pi}(\varphi_{mod}(i) + \varphi_{\nu}(i))\rfloor \% M$.
Here, $\lfloor x\rfloor$ denotes the largest integer less than or equal to $x$.
Normally, $M$ can be set to the order of $100$ for a balance between computational accuracy and speed.
With such grouping, group $G(k)$ contains all detection events with relative phase between $\phi_c+\frac{2\pi}{M}k$ and $\phi_c+\frac{2\pi}{M}(k+1)$, corresponding to an average phase of $\phi_c+\frac{2\pi}{M}(k+\frac{1}{2})$.
The event count from detector channels 0 and 1 are denoted as $C_{0}(k)$ and $C_{1}(k)$, respectively.
If $C_0(k)+C_1(k) > 0$, then group $G(k)$ is marked as valid, and the visibility is calculated as $V(k)=\frac{C_0(k)-C_1(k)}{C_0(k)+C_1(k)}$.
Subsequently, the error between the measured and the expected visibilities can be calculated as $\delta(\phi)=\Sigma_k [V(k) - sin(\phi+\frac{2\pi}{M}(k+\frac{1}{2}))]^2$.
$\phi$ is searched over $(0, 2\pi)$ to minimize the error $\delta(\phi)$, thus serves as the optimal estimation of $\phi_c$.


\end{document}